\documentclass[prd,aps,floats,nofootinbib]{revtex4}

\usepackage{epsfig}

\newcommand{\eqref}[1]{(\ref{#1})}

\begin{document}
%
%==============================================================================

%
% Definitions for this paper.
%

\def\e{\epsilon}
\def\apm{\alpha^\prime}

%
% Text of paper
%

\title{Black hole-black string phase transitions in thermal $1+1$ dimensional \\
supersymmetric Yang-Mills theory on a circle \\}

\author{Ofer Aharony}
\email{Ofer.Aharony@weizmann.ac.il}
\affiliation{Department of Particle Physics, Weizmann Institute of Science, Rehovot 76100, Israel}

\author{Joseph Marsano}
\email{marsano@fas.harvard.edu}
\affiliation{Jefferson Physical Laboratory, Harvard University, Cambridge MA 02138, USA}

\author{Shiraz Minwalla}
\email{minwalla@bose.harvard.edu}
\affiliation{Jefferson Physical Laboratory, Harvard University, Cambridge MA 02138, USA}

\author{Toby Wiseman}
\email{twiseman@fas.harvard.edu}
\affiliation{Jefferson Physical Laboratory, Harvard University, Cambridge MA 02138, USA}

\date{June 2004}

%==============================================================================
%
\begin{abstract}
%
%==============================================================================

  \vspace*{.3cm} {\hfill hep-th/0406210, WIS/17/04-JUNE-DPP } \vspace*{.3cm}

We review and extend earlier work that uses the AdS/CFT correspondence to
relate the black hole-black string transition of gravitational
theories on a circle to a phase transition in maximally
supersymmetric $1+1$ dimensional $SU(N)$ gauge theories at large
$N$, again compactified on a circle. We perform gravity calculations
to determine a likely phase diagram for the strongly coupled gauge
theory.
We then directly study the phase structure of the same gauge theory,
now at weak 't Hooft coupling. In the interesting temperature regime
for the phase transition, 
the $1+1$ dimensional theory reduces 
to a $0+1$ dimensional bosonic theory, which we solve using Monte Carlo
methods.  We find strong evidence that the weakly coupled gauge theory
also exhibits a black hole-black string like phase transition in the
large $N$ limit. We demonstrate that a simple Landau-Ginzburg like
model describes the behaviour near the phase transition remarkably
well. The weak coupling transition appears to be close to the cusp between
a first order and a second order transition.

%==============================================================================
%
\end{abstract}
%
%==============================================================================

\maketitle

%==============================================================================
%
\section{Introduction}
\label{sec:intro}
%
%==============================================================================

The gravitational theories dual to strongly coupled large $N$ 
Yang-Mills theories \cite{Maldacena:1998re,Gubser:1998bc,Witten:1998qj} sometimes exhibit particularly dramatic behaviours; they
undergo phase transitions involving the nucleation of black holes or
transitions between different black holes. Examples include the
Hawking-Page transition \cite{{Hawking:1983dh},{Witten:1998zw}} in AdS
spaces and the Gregory-Laflamme transition
\cite{Gregory_Laflamme1,Gregory_Laflamme3,Gregory_Laflamme4} in the
near-horizon geometry of D0-branes on a circle
\cite{{Susskind:1998vy},{Barbon:1998cr},{Li:1998jy},{Martinec:1998ja}},
the topic of this letter. These transitions have a dual description as
thermal phase transitions in strongly coupled Yang-Mills theories.

It is interesting to investigate how these phase transitions continue
to weaker Yang-Mills coupling\footnote{Similar continuations of strong
coupling results to weak coupling were recently discussed in
\cite{Fidkowski:2004fc}.}.  In the two examples described above,
the high temperature phase is (in the large $N$ limit)
sharply distinguished from its low
temperature counterpart by an order parameter. As a consequence, as
we decrease the coupling, the
line of phase transitions cannot end abruptly in $\lambda, T$ space
($\lambda=g_{YM}^2 N$ is the 't Hooft coupling and $T$ is the
temperature), and so it must either go to zero or infinite temperature
at a finite value of the coupling constant, 
or (as turns out to be the case in both of the
examples described above) it must continue all the way to weak
coupling.

The first example described above, the Hawking-Page transition, is
dual to a thermal deconfinement transition for the $3+1$ dimensional
${\cal N}=4$ supersymmetric Yang-Mills
theory on $S^3$. It turns out that this transition may indeed be
continued to weak coupling \cite{{Sundborg:1999ue},{Aharony:2003sx},
{next_paper}}
where it may be studied in detail with intriguing results.

In this letter we concentrate on the second example, the
Gregory-Laflamme transition
\cite{Gregory_Laflamme1,Gregory_Laflamme3,Gregory_Laflamme4}.
Following
\cite{{Susskind:1998vy},{Barbon:1998cr},{Li:1998jy},{Martinec:1998ja}},
we first review the dual interpretation of this transition as a
thermal phase transition in large $N$ $1+1$-dimensional maximally
supersymmetric Yang-Mills (SYM) theory on a circle of circumference
$L$, at strong coupling ($\lambda L^2 \gg 1$) \footnote{More
precisely, it turns out that the phase transition in this gauge theory
is dual to the Gregory-Laflamme transition 
of a collection of near extremal charged black holes in a space which is
asymptotically $R^{8,1}\times
S^1$. The transition may be described by a gauge theory, even though the
gauge theory is only dual to the near-horizon limit of the black holes, 
because the
unstable mode is localized purely within the near-horizon geometry,
as we will show below.}. 
The uniform black string corresponds to a phase in which
the eigenvalues of the holonomy of the gauge field around the spatial
circle are uniformly distributed on a unit circle in the complex
plane. The black hole corresponds to a phase in which these
eigenvalues are clumped about a particular point on the circle. We
then demonstrate that Yang-Mills theory on a circle at weak coupling
($\lambda L^2 \ll 1$) undergoes a phase transition distinguished by
the same order parameter. We conjecture that this phase transition is
the continuation to weak coupling of the black hole - black string
transition.

Our analysis of the thermal behaviour of maximally supersymmetric
$SU(N)$ Yang-Mills theory on a circle at small $\lambda L^2$ proceeds in
two stages. We first 
demonstrate that when $L^3 \lambda T \ll 1$ (and also $T^3 L \gg
\lambda$), the eigenvalues are sharply localized about a point on the
unit circle in the complex plane, in an arc of length $s \sim (L^3
\lambda T)^{1/4}$.  Continuing to higher temperatures, it is plausible
that at $L^3 \lambda T $ of order one these eigenvalues fill out the
circle and undergo a Gross-Witten-like \cite{Gross:1980he,Wadia:1979vk} `black
hole-black string' transition\footnote{In this paper what we mean by a
Gross-Witten transition is any transition in which the eigenvalue
distribution develops a gap; as discussed in \cite{Aharony:2003sx}
such a transition does not have to be continuous.}. However,
perturbation theory breaks down precisely at $T \sim 1 / \lambda
L^3$. In fact, at high temperatures the two-dimensional theory under
study effectively reduces to a one-dimensional Yang-Mills theory with
adjoint scalar fields, with 't Hooft coupling constant $\lambda T$,
compactified on a circle of circumference $L$. The effective coupling
of this one-dimensional system, $L^3 \lambda T$, climbs to unity
precisely when the eigenvalue dynamics gets interesting. Since we do
not know how to study this system analytically, the second stage of
our analysis involves a Monte Carlo simulation of this effective
strongly coupled $0+1$-dimensional bosonic system. We find good
evidence that there is indeed a sharp phase transition from a
localized to a uniform (smeared) eigenvalue distribution, at $T
\approx { 1.4 / \lambda L^3} $, and we analyze it in detail. To
conclude our note we present a Landau-Ginzburg model that reproduces
our detailed numerical results for the eigenvalue dynamics with remarkable
accuracy.

The work reported in this letter is part of a more comprehensive
investigation of the thermal properties of large $N$ Yang-Mills
theories on tori that will appear in a separate publication
\cite{torus}. In this note we will present the principle results that
have bearing on the Gregory-Laflamme transition. Section
\ref{sec:dual} contains an analysis of the gravitational system and
its translation to Yang-Mills theory. Our main results are in section
\ref{sec:weak}, where we analyze the Yang-Mills theory at weak
coupling. We end with a summary in section \ref{sec:summary}.  Two
appendices contain technical results which are used in section
\ref{sec:dual}.

%==============================================================================
%
\section{Holographic duality}
\label{sec:dual}
%
%==============================================================================

In this section we review and extend the arguments of
\cite{{Susskind:1998vy},{Barbon:1998cr},{Li:1998jy},{Martinec:1998ja}}
that relate the thermodynamics of strongly coupled $1+1$-dimensional
gauge theories in the 't Hooft large $N$ limit to the Gregory-Laflamme
transition. In gravity this transition is usually discussed as a function
of energy (mass), but in the gauge theory we can also discuss it as a
function of temperature (using the relation between the canonical and
micro-canonical ensembles), and we will use both languages interchangeably.
Consider a $1+1$ dimensional maximally supersymmetric Yang-Mills
theory at temperature $T$ on a circle of circumference $L$, in the 't
Hooft large $N$ limit with 't Hooft coupling $\lambda$. Define a
dimensionless coupling $\lambda' \equiv \lambda L^2$ and a dimensionless
temperature $t\equiv TL$. The Maldacena dual of this system is string theory
on the space that is obtained upon performing a circle identification
$\theta \equiv \theta + 2\pi$ on the familiar background
\cite{Itzhaki:1998dd} which is the near-horizon limit of 
an infinitely extended near extremal
$D1$-brane, with the string frame metric and the dilaton given by
\begin{eqnarray}
ds^2&=&\alpha'\left\{\frac{u^3}{\sqrt{d_1
	\lambda'}}\left[-\left(1-\frac{u_0^6}{u^6}\right)\frac{d\tau^2}{L^2}+ 
{d\theta^2\over {(2\pi)^2}} \right] +\frac{\sqrt{ d_1
\lambda'}}{u^3\left(1-\frac{u_0^6}{u^6}\right)}du^2+u^{-1}\sqrt{d_1\lambda'}d\Omega_7^2\right\},\cr
e^{\phi}&=&2\pi\frac{\lambda'}{N}\sqrt{\frac{d_1\lambda'}{u^6}},
\label{eq:sugraph}
\end{eqnarray}
where
\begin{equation}u_0^6=2^73\pi^5\epsilon  \left(\frac{\lambda'}{N}\right)^2,\qquad d_1=2^6\pi^3,
\label{eq:untdef}
\end{equation}
and the dimensionless energy, $\epsilon$, is given by ${\e }=EL$
($E$ is the ADM energy of the solution above extremality; note that we are
using a dimensionless $u$ coordinate as opposed to the conventions of
\cite{Itzhaki:1998dd}). There is also a non-trivial RR 3-form field
strength. The
entropy of the solution, as a function of energy, is given by
\begin{equation}S=\frac{2^{2/3}\pi^{5/6}N^{2/3}\epsilon^{2/3}}{3^{1/3}\lambda^{\prime
      1/6}}.
\label{eq:entropy}
\end{equation}
Using \eqref{eq:entropy}, one can obtain $u_0$ as a function of the
dimensionless temperature $t$,
\begin{equation}u_0^2={16 \pi^{{5/2}} \over 3}t \sqrt{\lambda'}.
\label{eq:uzert}
\end{equation}
 
Under what conditions are the stringy corrections to the
(non-supersymmetric) supergravity solution \eqref{eq:sugraph}
negligible?  In the neighbourhood of its horizon, \eqref{eq:sugraph}
is characterized by a single length scale $ l \sim \sqrt{\apm}
\lambda^{\prime {1 \over 4}} u_0^{-{1 \over 2}}$. Consequently,
$\alpha'$ corrections to \eqref{eq:sugraph} are negligible when $l \gg
\sqrt{\apm}$, i.e. when $t \ll \sqrt{\lambda'}$.  Winding modes (whose
mass in the neighbourhood of the horizon is $M_w\sim u_0^{3\over 2} /
\sqrt{\apm} \lambda^{\prime {1\over 4}}$) are negligible when $l M_w
\gg 1$, i.e. for $ t \gg 1 / \sqrt{\lambda'}$.  Thus, when $\lambda'
\gg 1$, the solution is valid over a large range of temperatures.

In order to obtain a supergravity description that is valid for $t$
lower than or of order $1/\sqrt{\lambda'}$, we T-dualize in the
$\theta$ direction to obtain the metric for a collection of
non-extremal D0-branes on a dual circle. In general, the distribution
of these 0-branes around the circle is dynamically determined.  At
large enough temperatures, they are uniformly distributed over the
circle, and the corresponding supergravity background is the T-dual of
\eqref{eq:sugraph},
\begin{eqnarray}
ds^2 & = & {\apm } \left( - {u^{3} } (1-{u_0^{6} \over u^{6}}) {d\tau^2
\over L^2\sqrt{\lambda' d_1}} + \sqrt{d_1\lambda'} u^{{-1 }} d
\Omega_{7}^2 +{\sqrt{d_1\lambda'} \over u^{3}
}\left[ {du^2 \over \left( 1-{u_0^{6} \over u^{6}} \right) } +
(2\pi)^2 d{\tilde \theta}^2 \right]\right), \cr
e^{\phi}&=&(2\pi)^2\frac{\lambda'}{N} \left( d_1\frac{\lambda'}{u^{6}}
\right)^{{3 \over 4}},
\label{eq:sugram}
\end{eqnarray}
where again ${\tilde \theta} \equiv {\tilde \theta} + 2\pi$ 
and there is a non-trivial RR 2-form
field strength. As above, $\alpha'$ corrections to the supergravity
solution \eqref{eq:sugram} (in the neighbourhood of the horizon) are
negligible for $t \ll \sqrt{\lambda'}$; winding modes about this
background are negligible provided $t \ll 1$.  In summary, stringy
corrections to the background \eqref{eq:sugram} are negligible at
large $\lambda'$ provided $ t \ll 1$.

Below we will also be interested in fluctuation modes about
\eqref{eq:sugram} that carry momentum about the circle. Such modes
will not excite string oscillators (and so will be well described by
supergravity) if and only if the proper length of the compact circle
(near the horizon)
is large in string units, i.e. for $t \ll 1/{\lambda'}^{{1 \over 6}}$.

%==============================================================================
%
\subsection{The uniform phase and the Gregory-Laflamme instability}
\label{sec:uniform}
%
%==============================================================================

Equation \eqref{eq:sugram} is the near-horizon geometry of a charged
black string in $R^{8,1} \times S^1$ (winding around the $S^1$); 
in this section, we demonstrate
that it develops a Gregory-Laflamme instability at a temperature $t\sim
1/\sqrt{\lambda'}$, which is well within the validity of the
supergravity approximation for large $\lambda'$. In order to
facilitate comparisons with previous analysis of the
Gregory-Laflamme transition we perform our analysis in the full
black brane solution (obtained by replacing $d_1 \lambda' / u^6 \to
(\alpha'/L^2)^2 + d_1 \lambda' / u^6$ in \eqref{eq:sugram}); as the
unstable mode turns out to be localized within the near-horizon
region, working directly in the near-horizon geometry \eqref{eq:sugram} will
yield identical results.

In order to analyze the instability it is useful to lift the
background \eqref{eq:sugram} to M theory; this is a useful trick for
simplifying the analysis, even though we are really only interested in
weakly coupled IIA backgrounds in which the circle of the eleventh
dimension is very small. Recall that $D0$-brane charge is
reinterpreted as momentum around the M theory circle; consequently the
M theory lift of \eqref{eq:sugram} is simply obtained from (the
near-horizon limit of) a toroidally compactified uncharged black
$2$-brane in M theory,
\begin{eqnarray}
ds^2 & = & - f(r) d\tau^2 + {dr^2\over f(r)} + r^2 d\Omega_{7}^2 + dy^2 + dx^2 ,\cr 
f(r) & = & 1 - {r_0^6 \over r^6},
\label{eq:metb}
\end{eqnarray}
by boosting along $x$, the M theory circle. This leads to
\begin{equation}
ds^2 = - d\tau^2 + dx^2 + \frac{1}{f(r)} dr^2 + r^2
d\Omega^2_{7} + dy^2 + \left( 1 - f(r) \right)
\left( \cosh{\beta} d\tau + \sinh{\beta} dx \right)^2 .
\label{eq:uniformboost}
\end{equation}
Here $y$ is a rescaled coordinate on our T-dual 
compact circle (proportional to 
${\tilde \theta}$ in \eqref{eq:sugram}) 
with period $y \equiv y + \tilde{L}$. One can
determine $\tilde{L}$ and the M theory parameters ($r_0$, the Planck
Mass, the radius $l_{11}$ of the M theory circle and the boost
rapidity $\beta$) such that the near-horizon limit of the background
\eqref{eq:uniformboost}, reduced back to 10-d along the $x$ direction, 
will be equal to
\eqref{eq:sugram}. In particular, it is easy to verify that $\tilde{L}
= (2\pi)^2 \alpha'/L$ and $u_0 = r_0 L / \alpha'$.

Fluctuations around the black brane solution
\eqref{eq:metb} include a Gregory-Laflamme zero
mode (carrying some momentum in the $y$ direction) when
$r_0/\tilde{L}$ is equal to some number $a(0)$, where one finds $a(0) \approx
0.37$.  As this mode is
independent of both $t$ and $x$, choosing a suitable gauge, it is
preserved by the boost (this argument was used in
\cite{Itzhaki:1998dd}).  Thus, we conclude that the fluctuations around
\eqref{eq:uniformboost}, and thus \eqref{eq:sugram}, also include a zero mode
at $u_0=(2\pi)^2 a(0)$. It is well-known that
in the solution \eqref{eq:metb} this zero mode bounds a region of 
instability \cite{Gregory_Laflamme1,Gregory_Laflamme3,Gregory_Laflamme4},
and we will see below that this is true also for the near-horizon limit of 
the boosted solution
\eqref{eq:sugram}, which becomes unstable when $u_0 < (2\pi)^2 a(0)$.
Hence, in the gauge theory, the phase corresponding to the uniform black
string becomes unstable below
a critical temperature
\begin{equation}t_{GL}={3 \over 4 \sqrt{\pi} }
    \frac{(2\pi a(0))^2}{\sqrt{\lambda'} }.
\label{eq:tcrit}
\end{equation}

In the rest of this section we explicitly demonstrate that the
fluctuation spectrum of \eqref{eq:uniformboost} includes an
instability for $u_0 < (2\pi)^2 a(0)$ (and thus, in the dual theory,
for temperatures lower than $t_{GL}$), and verify that this unstable
fluctuation is supported purely within the near-horizon geometry.  The
perturbation of interest takes the form
\begin{equation}
h_{\mu\nu}(\tau,r,y) = e^{\tilde{\Omega} \tau } \Lambda_{\mu}^{\mu'} \Lambda_{\nu}^{\nu'} \left[ 
\cos{\tilde{k} y} b_{\mu'\nu'}+\sin{\tilde{k} y} c_{\mu'\nu'}
\right],
\end{equation}
where $\Lambda$ is the Lorentz transformation corresponding to the boost,
\begin{eqnarray}
\tilde{\Omega} & = & \frac{\Omega}{\cosh{\beta}}, \cr
\tilde{k}^2    & = & k^2 + \Omega^2 \tanh^2{\beta},  
\label{eq:relations}
\end{eqnarray}
and the nonzero components of $b,c$ are
\begin{eqnarray}
b_{\tau \tau} & = & h_{\tau}(r) ,\; \;
b_{rr} = h_r(r) ,\; \; b_{\tau r} = \Omega h_v(r), \cr
b_{yy} & = & \cosh^2{\alpha} h_y(r),\;\;
b_{xx} = - \sinh^2{\alpha} h_y(r),\; \;
b_{rx} = k \sinh{\alpha} h_v(r), \cr
c_{ry} & = & k \cosh{\alpha} h_v(r), \; \;
c_{xy} = - k \sinh{\alpha}\cosh{\alpha} h_y(r),
\label{eq:tensor}
\end{eqnarray}
where $\alpha$ is defined by $k \sinh{\alpha} = \Omega
\tanh{\beta}$. Note that the perturbation has no functional dependence
on the $x$ coordinate so its reduction to IIA supergravity is
straightforward, although it does involve the type IIA RR 1-form and
dilaton as well as the metric.

It follows from the linearized Einstein equations that $h_y$ obeys a
simple second order eigenvalue equation in $\Omega^2$,
\begin{eqnarray}
h_y''(r) & + & p(r) h_y'(r) + q(r) h_y(r) = \Omega^2 w(r) h_y(r), \cr \cr
p(r) & = & \frac{1}{r} \left( 1 + \frac{6}{f(r)} - \frac{16 \, k^2 r^2}{k^2 r^2 + 21 {r_0^6\over r^6}} \right),  \cr
q(r) & = & \frac{1}{r^2} \left( - \frac{k^2 r^2}{f(r)} \; \frac{k^2 r^2 - 27 {r_0^6\over r^6}}{k^2 r^2 + 21 {r_0^6\over r^6}} \right), \cr
w(r) & = & \frac{1}{f(r)^2}.
\label{eq:eqnhx}
\end{eqnarray}
Once $h_y$ is determined, $h_{\tau}$, $h_r$, and $h_v$ 
may be simply derived from
it. The boundary conditions are as for the original Gregory-Laflamme
analysis \cite{Gregory_Laflamme1}.

We see that \eqref{eq:eqnhx} is completely independent of the boost parameter
$\beta$. Hence, from our knowledge of the unboosted problem, we know that for
$k < k_c = 2 \pi a(0)/r_0$ there exist physical modes with real positive
$\Omega$ that lead to dynamical instability of \eqref{eq:uniformboost}
for all $\beta$.  Of
course, the precise nature of this unstable mode does depend on the
boost; note that the instability exponent $\tilde{\Omega}$, the
wavelength $\tilde{k}$, and the tensor structure -- through $\alpha$
-- are all functions of $\beta$.

As for the unboosted case, the unstable modes decay away from the
horizon as $h_{\mu\nu} \sim e^{- 2 \pi a(\Omega) r/r_0}$, where
$a(\Omega)$ is a number of order unity, and $0 < a(\Omega) <
a(0)$. The translation to \eqref{eq:sugram} takes $r/r_0$ to
$u/u_0$. Consequently, the unstable mode is a normalizable dynamical
degree of freedom even strictly in the near-horizon limit
\eqref{eq:sugram}
\footnote{The thermodynamics of the IIA uniform smeared near-extremal
D0-brane gravity solution \eqref{eq:sugram} is the same as that of the
T-dual D1-brane background \eqref{eq:sugraph}. As we have shown, the
smeared D0 solution is unstable up to extremality (for large enough
circle radius), whereas the T-dual D1 solution is stable away from
extremality \cite{Gubser_brane}. The Gubser-Mitra conjecture
\cite{Gubser_Mitra1,Gubser_Mitra2,Reall,Gregory:2001bd,Hubeny_Rangamani,Hartnoll:2004kz}
links dynamical and local thermodynamic stability. It is easy to show
that if for the D0 solution one allows both the charge and the mass to
vary when computing the thermodynamic stability as in
\cite{Gubser_Mitra2}, then this background is indeed predicted to be
thermodynamically unstable up to extremality, whereas the D1 solution
with fixed charge is stable. Physically, allowing the D0 charge to
vary is sensible as the D0 charge can be defined as an integral over a
local density on the torus, just as the mass density is local, whereas
for the D1 solution a charge density cannot be constructed, and thus
we think of it as a global quantity, and hence fixed. Essentially we
should allow the D0 mass and the charge to be able to redistribute
themselves over the torus when computing the thermodynamic
stability. See also the recent \cite{Bostock:2004mg}. }.

%==============================================================================
%
\subsection{Other phases: Localized and non-uniform}
\label{sec:otherphases}
%
%==============================================================================

Following
\cite{{Susskind:1998vy},{Barbon:1998cr},{Li:1998jy},{Martinec:1998ja}}
we have argued that strongly coupled maximally supersymmetric $1+1$
dimensional Yang-Mills theory on a circle undergoes a phase transition
at a temperature of order $t \sim 1 / \sqrt{\lambda'}$, since the high
temperature phase becomes unstable there. We now wish to investigate how the
system behaves at lower temperatures.

As we have seen, the AdS/CFT correspondence maps (the near horizon limit of)
near extremal charged
black solutions of IIA supergravity on $R^{8, 1} \times S^1$ to phases
of maximally supersymmetric Yang-Mills theory on $S^1$. The
thermodynamic properties of a given Yang-Mills phase are easily
obtained from the Bekenstein-Hawking thermodynamics of the
corresponding charged black solution. Furthermore, we have seen in the
previous section that we may generate near extremal charged solutions
from uncharged solutions (via an M-theory lift-boost-reduction)\footnote{
Such a procedure for ``charging up'' solutions was previously used in
\cite{{Hassan:1991mq},{Harmark_Obers}}.}. Hence, a
complete understanding of uncharged solutions on $R^{8,1}\times S^1$
would fix the phase structure of the Yang-Mills theory under
study (at strong coupling, where the supergravity approximation is
valid). In this section, we review what is known about these uncharged
solutions for the theory on a circle.

%==============================================================================
%
\subsubsection{Uncharged solutions}
\label{sec:uncharged}
%
%==============================================================================

So far three branches of uncharged solutions on spaces with an
asymptotic circle have been found. These are the uniform black string
(discussed in the last section), the localized black hole, and the
non-uniform string, which are distinguished asymptotically by two
gravitational charges
\cite{Townsend:2001rg,Traschen:2003jm,Shiromizu:2003gc,Kol:2003if,Harmark:2003dg}
\footnote{With the $R^{d-1} \times S^1$ asymptotics there are two
gravitational charges, and interestingly the known solutions appear to
be uniquely distinguished by them \cite{Harmark:2003eg}, unlike
general higher dimensional black holes which violate uniqueness in
terms of the asymptotic charges \cite{Emparan_Reall2}.}.

As outlined above, uniform strings on a circle are stable for masses
larger than the mass of the string which possesses the
Gregory-Laflamme zero mode. Below this mass they are dynamically
unstable. The known non-uniform string solutions live on a branch
which emerges from the uniform string branch at the marginal
point. For the dimensions of interest, near the marginal point the
non-uniform solutions have larger mass than the unstable uniform
strings, and lower entropy than a stable uniform string with the same
mass \cite{Gubser,Sorkin:2004qq}. In the 6-d case these solutions have
been numerically constructed away from the marginal point, and these
properties appear to continue to the end of the branch where the
minimal sphere of the horizon pinches off \cite{Wiseman3}. However,
the 10-d solutions have not been constructed so far.

Localized black holes are believed to be stable at low energies.
These solutions have recently been constructed numerically in 5-d and
6-d \cite{Kudoh:2003ki,Sorkin:2003ka}, and they can also be constructed
in perturbation theory,
expanding in the black hole mass
\cite{Harmark:2003yz,Gorbonos:2004uc} \footnote{Analytic constructions are
restricted to 4-d \cite{Myers}, where there are no black string
solutions and hence no Gregory-Laflamme dynamics.}.

Kol has suggested that the black holes grow until they cannot `fit' in the
circle, and then they join the end of the non-uniform string 
branch via a cone-like
topology changing solution \cite{Kol1} (see also \cite{Harmark_Obers}). 
Evidence supporting this
elegant picture has recently emerged from the numerical solutions
\cite{Wiseman4,Kol_Wiseman,Toby_Kudoh}. As pointed out by Kol, this
picture seems rather natural when considering a Gross-Witten like
eigenvalue transition.

%==============================================================================
%
\subsubsection{Near-extremal charged solutions}
\label{sec:nearext}
%
%==============================================================================

As mentioned above, if we knew the 10-d uncharged solutions on a
circle, we would then simply be able to translate them into
near-extremal charged solutions which are relevant for the gauge
theory, and to plot a phase diagram for the near extremal solutions.
In appendix A we explicitly show how this would be done, by `charging'
an uncharged solution via the M-theory lift-boost-reduce procedure,
and then examining the near extremal limit. This process would
determine which solution is thermodynamically favoured at a given
energy or temperature.

In Appendix B we take the first steps in implementing this
programme. We use Gubser's perturbative construction \cite{Gubser} to
repeat Sorkin's calculation \cite{Sorkin:2004qq} of the 10-d
non-uniform strings. We then apply the M-theory charging procedure of
Appendix A to determine the thermodynamic properties of near-extremal
weakly non-uniform solutions. Our results demonstrate that, within the
perturbative approach outlined above, near extremal non-uniform
strings are thermodynamically disfavoured over the stable uniform
strings, just like their uncharged counterparts. In particular, in the
canonical ensemble, non-uniform solutions have a less favourable free
energy than the uniform strings of the same temperature.

Unfortunately, full numerical constructions for all types of uncharged
solutions in 10-d (that would extend the results of Appendix B to
increasingly non-uniform strings, and would supply similar results for
localized black holes) are not yet available. The equivalent
constructions currently exist only in 6-d. However, we think it
plausible that Kol's picture is also true for the 10-d uncharged
solutions (and for other dimensions obeying $d \leq 13$), and further
that the thermodynamic ordering of the three phases is preserved by
boosting. As evidence for the second part of this statement, we
explicitly verify in Appendix A that the boost does preserve the order
of entropy as a function of mass of the three solutions for the known
6-d solutions.

\begin{figure}
\epsfig{file=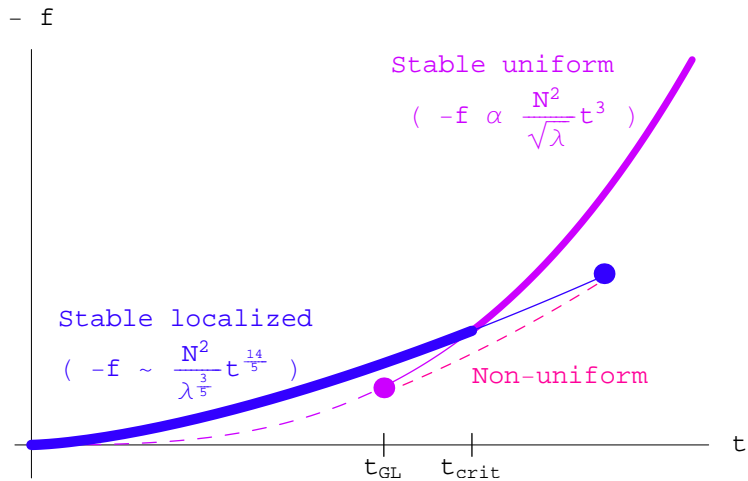,width=4in}
\caption{An educated guess for the free energy as a function of
temperature for black holes, uniform black strings, and non-uniform
black strings in 10 dimensions. Appendix B shows that the non-uniform
solutions have less favoured free energy than the uniform solutions
near $t_{GL}$. Analogy with $d = 6$ indicates that the non-uniform and
localized phases may meet at a topology changing solution. Dashed lines
represent phases conjectured to be unstable. 
\label{fig:phase}
}
\end{figure}

A conjectured plot of the free energy as a function of temperature for strongly
coupled Yang-Mills theory, that is consistent with the above
information, is shown in figure \ref{fig:phase}. As a function of the
temperature, the system in the figure undergoes a single first-order phase
transition, from a low temperature black hole phase to a high
temperature uniform black string phase, at a temperature above the
Gregory-Laflamme critical temperature \eqref{eq:tcrit}.

We emphasize that this figure is only a guess, based largely on the results of
section \ref{sec:uniform}, Kol's picture and the calculation in
Appendix B. It will be possible to verify this figure (and the ensuing
phase diagram)
once numerical (or other) data on the
properties of black holes and black strings in 10 dimensional gravity on
a circle becomes available.  

%==============================================================================
%
\subsection{Wilson loops as order parameters}
\label{sec:polyakov}
%
%==============================================================================

Euclidean $1+1$-dimensional $SU(N)$ 
SYM on a torus has two non-contractible Wilson loops,
namely the Wilson loops running around the time ($\tau$) and space
($x$) circles. Consider the spatial Wilson loop,
\begin{equation}
P_x = Pe^{ i \oint dx A_x }.
\end{equation}
According to the usual rules of T-duality, when we consider this
theory as coming from D1-branes on a circle, the phases of the
eigenvalues of $P_x$ represent the positions of T-dual D0-branes on
the dual circle (in a similar fashion, the eigenvalue distributions of
the 8 SYM adjoint scalar zero modes represent the transverse positions
of the D0-branes).  Hence, the 3 potential phases discussed above -- uniform,
non-uniform and localized -- are expected to be distinguished by the
eigenvalue distribution of $P_x$ (which is continuous in the large $N$
limit).  The uniform string corresponds to a uniform eigenvalue
distribution. The non-uniform string maps to a non-uniform eigenvalue
distribution that breaks translational invariance but is nowhere zero
on the unit circle in the complex plane. The black hole is expected to
be characterized by an eigenvalue distribution that is sharply
localized on the unit circle (a distribution that is strictly zero
outside an arc on the circle).

The temporal Wilson loop $P_{\tau}$ is expected to have a localized eigenvalue
distribution for all of these phases, corresponding to the presence of
a horizon \cite{Witten:1998zw,Aharony:2003sx}.

%==============================================================================
%
\section{Supersymmetric Yang-Mills theory at weak coupling}
\label{sec:weak}
%
%==============================================================================

We now turn to the study of the thermodynamics of the $1+1$-dimensional
maximally supersymmetric
$SU(N)$ Yang-Mills theory at weak coupling.  The Euclidean action for the SYM
theory is
\begin{equation}
S = \frac{1}{4g_{YM}^2}\int d\tau dx \; \mathrm{Tr} \left( F_{\alpha \beta}^2 
+ 2\sum_I D^{\alpha}\phi^{I} D_{\alpha}\phi^{I}  
-\sum_{I,J}\left[\phi^{I},\phi^{J}\right]^2 + \mathrm{fermions} \right),
\label{eq:ymaction}
\end{equation}
where $\phi^I$ are 8 adjoint scalars, $x$ is periodic with period $L$, 
and $\tau$, the Euclidean time,
is periodic with period $\beta=L/t$. Both scalars and fermions are taken to
be periodic on the spatial circle $x$. As usual, the fermions are
anti-periodic in $\tau$, distinguishing the temporal and spatial
directions.

%==============================================================================
%
\subsection{Zero mode reduction}
\label{sec:zeromode}
%
%==============================================================================

The first approximation to this $1+1$-dimensional theory is to reduce to
zero modes on the 2-torus. Kaluza-Klein (KK) modes on both cycles of the torus
are weakly coupled and can be integrated out within the window
of opportunity:
\begin{eqnarray}
\mathrm{temporal\;KK} & \qquad \qquad \mathrm{zero\;mode} \qquad \qquad & \mathrm{spatial\;KK} \cr
\mathrm{strongly\;coupled}  & \mathrm{reduction} &  \mathrm{strongly\;coupled}  \cr
t < \lambda^{\prime 1/3} \quad & \lambda^{\prime 1/3} < t < \frac{1}{\lambda'} & \quad t > \frac{1}{\lambda'}
\end{eqnarray}
In the intermediate 
regime, the theory is well described by a simple bosonic $SU(N)$
matrix integral, which has previously been studied in
\cite{Hotta:1998en},
\begin{equation}
Z = \int d\psi e^{ \frac{N}{4} \mathrm{Tr} \left( \sum_{A,B} \left[ \psi^{A}, \psi^{B} \right]^2 \right) },
\label{eq:matrixintegral}
\end{equation}
where now the 10 traceless adjoint matrices $\psi^A$ are composed of
the zero modes of the 8 traceless scalars and the gauge potential as:
\begin{equation}
\{ \psi^0, \psi^1, \psi^I \} \equiv \frac{1}{\beta\left(\lambda' t\right)^{1/4}} 
\int d\tau dx \{ A_{\tau}, A_{x}, \phi^I \}.
\end{equation}
Note that no fermion zero modes survive the reduction as they are
anti-periodic on the temporal circle.

\begin{figure}
\centerline{
\epsfig{file=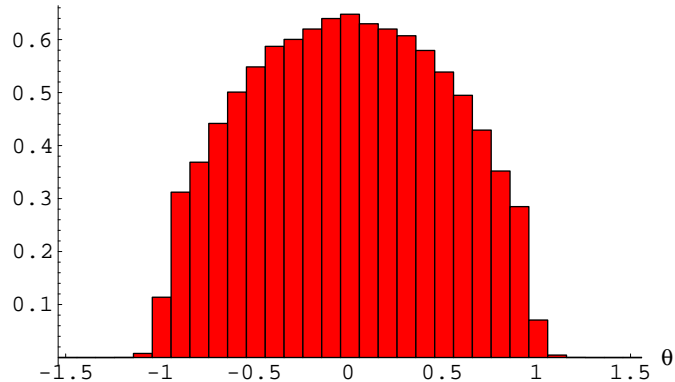,width=3.5in}
}
\caption{Monte Carlo determination of the saddle point eigenvalue
distribution of $\psi^A$ for the matrix integral \eqref{eq:matrixintegral}
with $N = 20$.
\label{fig:matrix}
}
\end{figure}

This theory has a classical moduli space along which the matrices
$\psi^A$, nine of which correspond 
to the spatial configuration of the D0-branes, are
diagonal. While this space is noncompact, the 1-loop effective
potential obtained by integrating out the off-diagonal modes in
\eqref{eq:matrixintegral} is
attractive; this ensures convergence of the moduli space integral.
However, this perturbative calculation is only valid when the
eigenvalues of $\psi^I$ are spread out over a distance scale that is
much larger than unity. At shorter separations, the off-diagonal modes
become light and hence strongly coupled.  It turns out that strong
coupling effects stabilize the eigenvalue distribution in a saddle
point that is sharply localized, with a characteristic 
scale of order unity. Note that
this scale is independent of $N$, as follows from 't Hooft scaling. We
have verified this claim by a Monte-Carlo integration, finding
$\langle \frac{1}{N} \mathrm{Tr}\left(\psi^A \psi^A \right) \rangle
\simeq 2.5$ in the large $N$ limit (this computation was first
performed by \cite{Hotta:1998en}; see also \cite{Krauth:1999qw}).  We
plot the eigenvalue distribution in figure \ref{fig:matrix}, which
shows that it is compactly supported. The spread of the eigenvalues of
the $\psi$'s is of order one, meaning that the spread of the
eigenvalues of the original matrices $A_{\tau}, A_x$ and $\phi^I$ is
of order $(\lambda' t)^{1/4} / L$.

%================================================\==============================
%
\subsection{High temperature reduction}
\label{sec:bosonic}
%
%==============================================================================

As we have argued in the previous subsection, the $1+1$-dimensional theory
under consideration reduces to a zero mode integral in the range
$\lambda^{\prime 1/3} \ll t \ll 1/\lambda'$.  It follows that
the $D0$ branes are clumped in a spherically symmetric distribution in
this range.  This may be thought of as the weak coupling analog of the
approximate spherical symmetry displayed by low temperature D0-brane
gravity solutions on a circle.

As the temperature is increased to $\lambda' t\sim 1$, the width of
the zero brane distribution approaches unity and the zero branes begin
to spread out over the circle. In the SYM theory, the spread of the
eigenvalue distribution of $A_x$ approaches the periodicity of this
variable (arising from large gauge transformations) which is
$2\pi/L$. It is natural to guess that the eigenvalue distribution
undergoes a Gross-Witten like unclumping transition at a temperature
of order $t \sim 1 / \lambda'$. Unfortunately, the spatial KK modes
become strongly coupled exactly around this temperature; consequently,
this guess is difficult to verify analytically. The spatial KK modes
are essential to the dynamics of the system at this temperature; they
are required in order to correctly reproduce the compactness of the
$SU(N)$ gauge group (the periodicity of $A_x$), and certainly cannot
be ignored.  Note, however, that temporal KK modes are utterly
negligible at this temperature. Consequently, the conjectured phase
transition should be well described by a $0+1$ dimensional bosonic gauge
theory obtained by dimensional reduction along the temporal
circle\footnote{Strictly speaking the system is well described by a
$0+1$ dimensional bosonic gauge theory with an inbuilt UV cutoff at
scale $\sim T$. This cutoff regulates the one-loop contribution to the
cosmological constant (the only divergent graph in the theory)
yielding a term in $\ln Z$ that scales like $N^2 \lambda L^3 T$. This
easily calculable term is the contribution of free high energy partons
to our system.  It dominates the partition function in the regime $t
\gg 1$ and $\lambda' \ll 1$ but makes no contribution to the
eigenvalue potential, and so is of limited relevance to the analysis
in this paper. See \cite{Gross:1981br} for similar remarks in the
context of 4 dimensional gauge theories. We thank A. Hashimoto for
discussions on this issue.}.  Again, the fermionic modes are
anti-periodic around the temporal circle, and so may be ignored. This
theory can be numerically solved using standard lattice Monte-Carlo
methods. Being a one-dimensional theory, the continuum limit is rather
simple to obtain, with relatively few lattice points being required
(even using only 5 points gives rather accurate results, with very
small corrections upon lattice refinement). The gauge dynamics in this
theory are trivial and can be removed, up to the spatial Wilson loop,
by choosing a gauge in which all the lattice link variables are
diagonal and equal -- of course, care must then be taken to implement
the measure correctly.

\begin{figure}
\epsfig{file=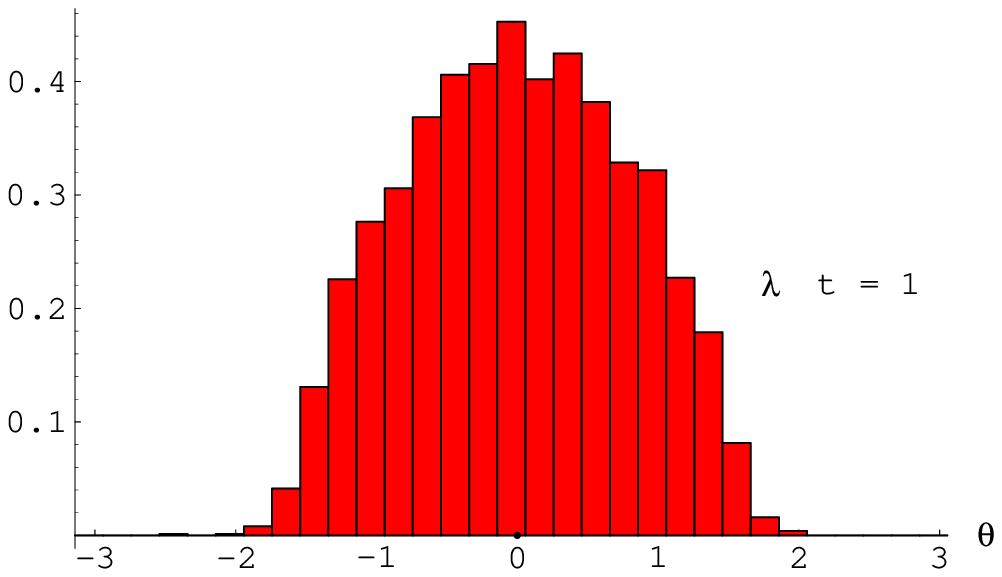,width=3.5in}
\epsfig{file=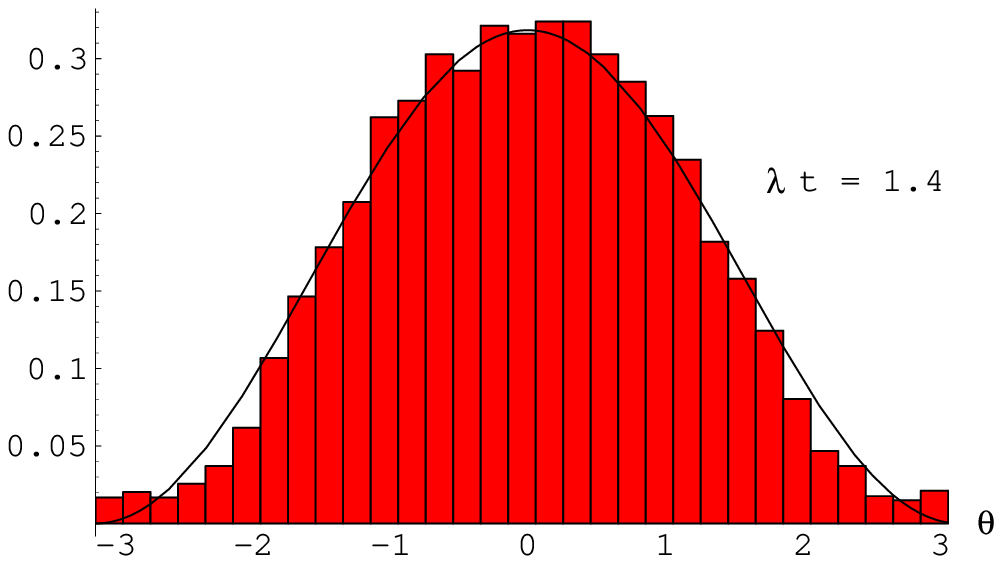,width=3.5in}
\epsfig{file=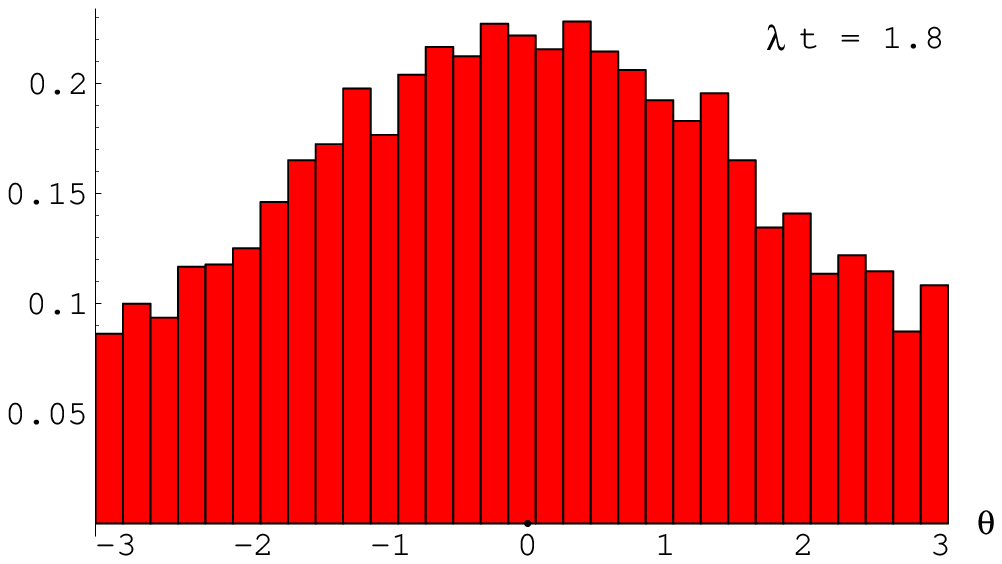,width=3.5in}
\caption{Eigenvalue distribution of $P_x$ for various values of
$\lambda' t$ with $N=12$. A phase transition appears to occur for
$\lambda' t \simeq 1.4$, and for this value we compare with $(1 +
\cos{\theta})/(2 \pi)$, the transition distribution for our $b = 0$
Landau-Ginzburg model discussed in section \ref{sec:LGanal}.
\label{fig:polyakov}
}
\end{figure}

The results of this simulation are indeed compatible with a Gross-Witten
like phase transition at $\lambda' t \simeq 1.4$ at large $N$. In figure
\ref{fig:polyakov} we show some typical eigenvalue distributions for
the spatial Wilson loop, for couplings below, around and above the
transition temperature. We clearly see that below the transition the
distribution has compact support, while above it is smeared over the
full angular period. 

\begin{figure*}
\epsfig{file=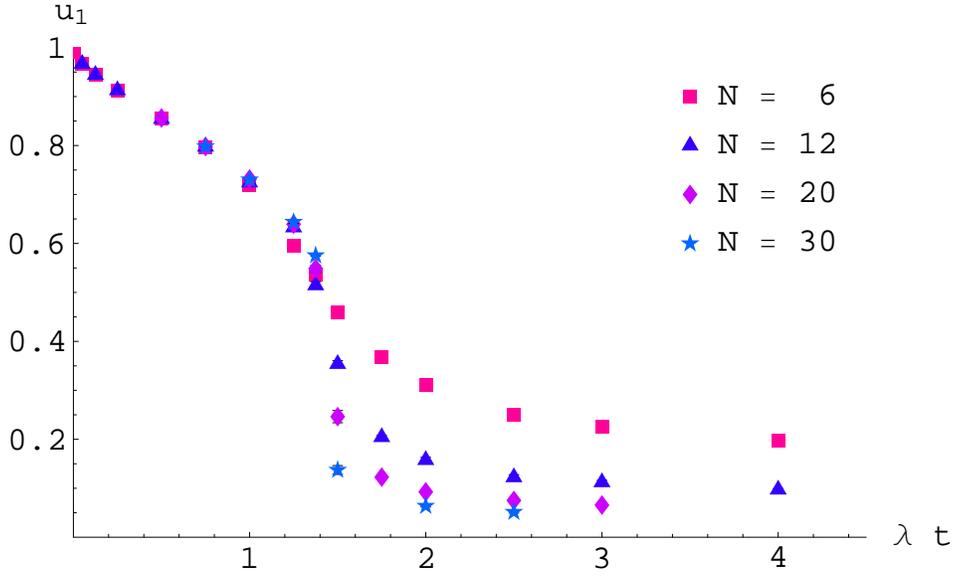,width=5in}
\caption{The first Fourier mode $u_1$ of the eigenvalue distribution
  of $P_x$ as a function of $\lambda' t$, for various values of
  $N$. To the left of the transition at $\lambda' t \simeq 1.4$, the
  values remain approximately invariant as $N$ increases.  To the
  right, $u_1$ decreases consistent with going to zero in the large
  $N$ limit as $1/N$. Statistical error bars are smaller than the plot
  symbols.
\label{fig:u1}
}
\end{figure*}

\begin{figure}
\epsfig{file=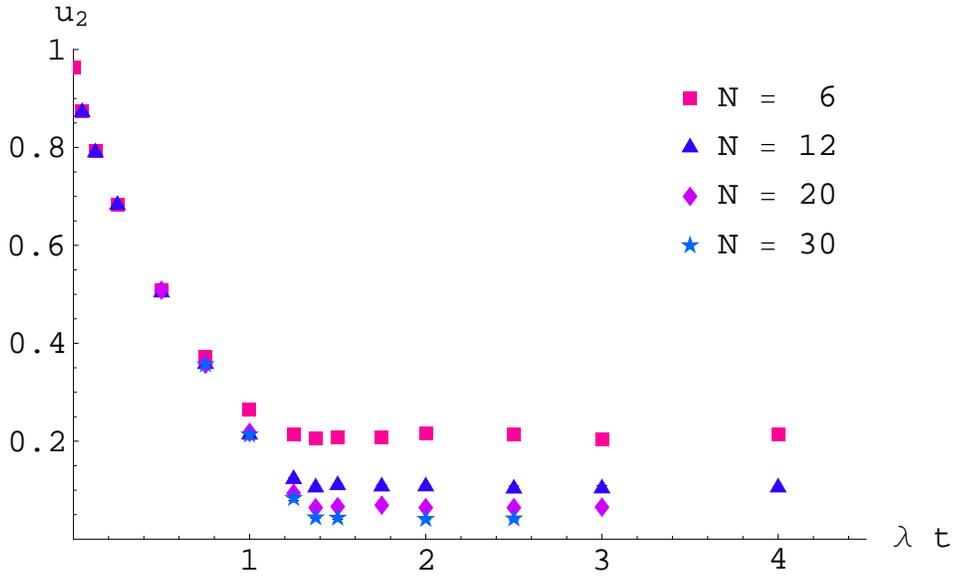,width=5in}
\caption{The second Fourier mode $u_2$ of the eigenvalue distribution
  of $P_x$ as a function of $\lambda' t$ for various values of
  $N$. For large $N$ the gradient appears to become discontinuous at
  the transition.
\label{fig:u2}
}
\end{figure}

\begin{figure}
\centerline{\epsfig{file=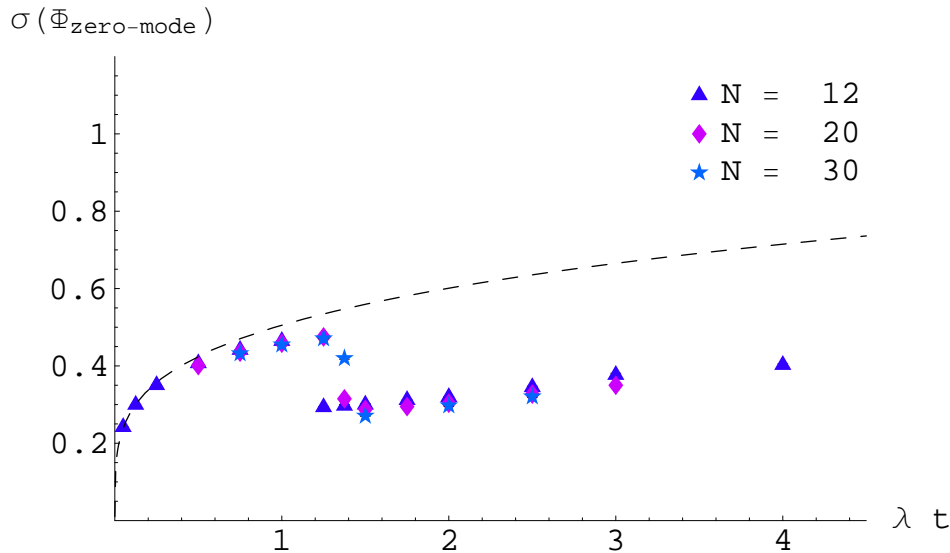,width=5in}}
\caption{The scalar zero mode eigenvalue distribution width as a
  function of $\lambda' t$, for various values of $N$. The dotted line
  gives the theoretical result from the $0+0$ dimensional matrix
  integral, which (as expected) reproduces the behaviour at small
  $\lambda' t$.
\label{fig:KK}
}
\end{figure}

As is apparent from figure \ref{fig:polyakov}, for $N=12$ the system
undergoes a smooth transition between the localized phase and the
non-uniform phase. However, this transition gets sharper as $N$ is
increased. This is evident by looking at the first two Fourier modes
of the eigenvalue distribution, $u_1$ and $u_2$ (defined by
$u_n=\mid\frac{1}{N}\sum_iz_{(i)}^n\mid={1\over
N}\mid\mathrm{tr}(P_x^n)\mid$, where $z_{(i)}$ denote the
eigenvalues of $P_x$), shown in figures \ref{fig:u1} and
\ref{fig:u2} as a function of $\lambda' t$ for various values of
$N$.  Note that the plot of $u_1$ versus $\lambda' t$ appears to
develop a sharp jump (from $u_1\approx 1/2$ to $u_1=0$ at $\lambda'
t\approx 1.4$) in the limit $N\rightarrow\infty$.  These results
strongly suggest the existence of a sharp phase transition at
$\lambda' t \approx 1.4$ and large $N$. Recall that, as discussed in
section \ref{sec:polyakov}, the $u_n$ are the expected order
parameters for the transition, and they should vanish in a ``uniform
black string''-like phase.  Note that the scalar zero mode also
responds sharply at the transition point (figure \ref{fig:KK}).

Interestingly, at the transition point $u_1 \simeq 0.5$ and the higher
Fourier modes are small. Thus it seems likely that $u_1$ is the
dominant order parameter for the transition, so that the first Fourier
component of the eigenvalue distribution gives the most relevant 
low mass mode near the transition.

%==============================================================================
%
\subsection{Landau-Ginzburg Analysis}
\label{sec:LGanal}
%
%==============================================================================

As we have noted above, the order parameter of our system appears to
jump discontinuously at $\lambda' t \simeq 1.4$ \footnote{Thus our
results disagree with the `mean field' prediction of a single
continuous phase transition in the same theory, 
reported in \cite{Kabat:1999hp}.}. It is
tempting to interpret this observation as evidence for a first order
phase transition.  However, it is also possible that, instead, the
system undergoes two continuous phase transitions, the first at $u_1
\sim 0.5$ and at $\lambda' t \approx 1.4$, and the second at $u_1 =0$
and $\lambda' t$ a little larger than $1.4$.  This possibility is not as
outlandish as it might first seem, as we now explain.

\begin{figure}
\centerline{\epsfig{file=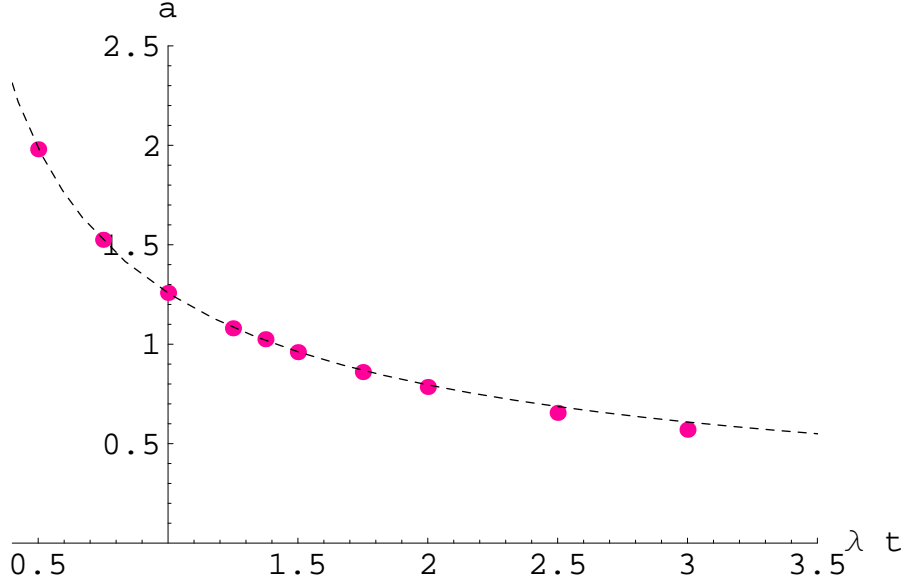,width=5in}}
\caption{
\label{fig:a}
The data points show the values of $a(\lambda' t)$ obtained by fitting
the matrix model with $b = 0$ to the $0+1$-dimensional lattice data.
The value of $a$ is determined by fitting for $u_1$ with $N = 6$. Note
that the behaviour of $a$ appears smooth across the transition point,
indicating that the matrix model correctly reproduces the transition
behaviour of $P_x$. The dashed line is the curve $1.3/(\lambda'
t)^{2/3}$, which appears to give a reasonable fit to the data points
over this range.  }
\end{figure}

\begin{figure}
\centerline{\epsfig{file=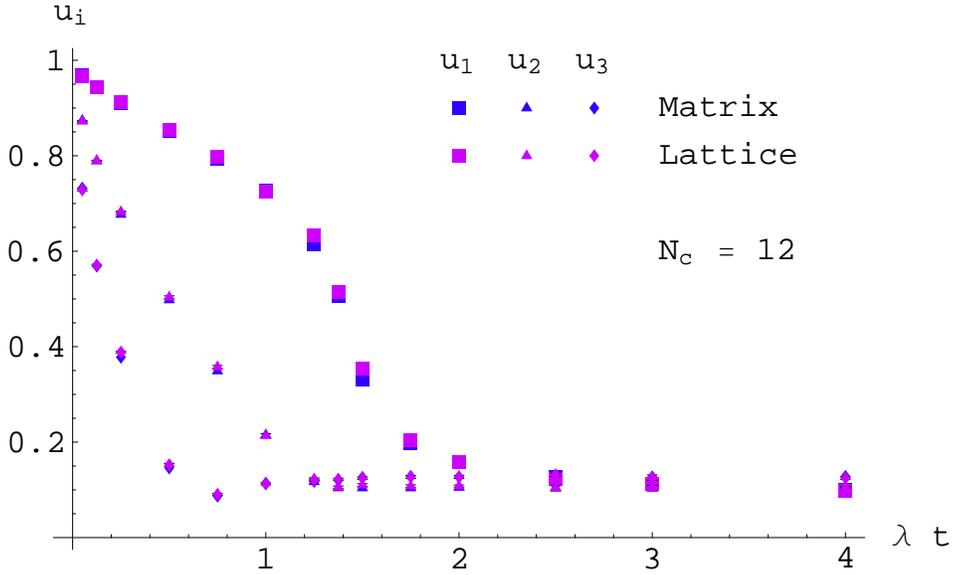,width=5in}}
\caption{Comparison of $0+1$-dimensional Monte Carlo data and
Landau-Ginzburg results for the matrix model with $b=0$, with $a$
determined as in the previous figure.  We clearly see excellent
agreement for the various moments $u_{1,2,3}$ for $N = 12$ as plotted
here.  A similar good agreement is observed for $N = 20,30$.
\label{fig:LGfits}
}
\end{figure}

We have attempted to fit the data from our Monte Carlo simulation to
the predictions from the following Landau-Ginzburg model, with $a$ and
$b$ smooth functions of $\lambda' t$,
\begin{equation}
Z_{MM} = \int dU e^{ a \mid \mathrm{Tr}(U) \mid^2 + 
b \mid \mathrm{Tr}(U) \mid^4 / N^2}.
\label{eq:lg}
\end{equation}
Identifying $U$ with $P_x$ (so that $u_n = {1\over N}
|\mathrm{Tr}(U^n)|$) the surprise is firstly that this works
remarkably well, and secondly that $b$ is small.  Let us set $b$ to
zero. In this case the model \eqref{eq:lg} exhibits a weakly first
order phase transition at $a=1$
\cite{{Sundborg:1999ue},{Aharony:2003sx}}. Figure \ref{fig:a} shows
the value of $a$ as a function of $\lambda' t$ found by fitting the values of
$u_1$ measured from the $N = 6$ lattice simulations to the values
coming from the matrix model for $N = 6$ (with $b=0$). Note that this curve
continues smoothly through the transition. For this parameterization
of $a$, in figure \ref{fig:LGfits} we then plot the predictions for
$u_1$ and $u_2$ from the model \eqref{eq:lg}, now with $N=12$, compared
to the actual data (also with $N = 12$). We find excellent agreement
within the error bars of the lattice data, and in we fact find a similar
agreement also for $N = 20, 30$.

It is striking that our Landau-Ginzburg model fits the Monte Carlo data so
well for $b \approx 0$. In order to explain the significance of this
observation, we will now digress to review the solution of the model
\eqref{eq:lg} at large $N$. As explained in section 6 of
\cite{Aharony:2003sx}, this model can be exactly solved in terms of
the Gross-Witten matrix integral. When $b$ is small and negative, this
model undergoes a single first order phase transition from a black
hole like clumped eigenvalue phase for $a> 1-|b|/4$ to a uniform
eigenvalue distribution for $a < 1-|b|/4$. The order parameter $u_1$
jumps discontinuously from $u_1 \sim {1\over 2}$ to $u_1 =0$ on going
through this transition.  When $b$ is small and positive, the system
undergoes two continuous phase transitions : from a clumped eigenvalue
phase (for $a>1+b/4$) to a non-uniform string phase (for $1<a<1+b/4$)
to a uniform string phase (for $a<1$). As $a$ varies between 1 and
$1+b/4$, the order parameter $u_1$ evolves continuously from ${1 \over
2}$ to zero. This continuous variation approaches a discontinuous
jump as $b\rightarrow 0$.

Our numerics are unable to distinguish very small values of $b$,
either positive or negative, from $b=0$ (hence a single first order
phase transition from two closely separated continuous
transitions). Our system clearly lies very near the cusp that
separates these two possibilities. It is even possible that $b$ is
exactly zero in the weak coupling limit for some good reason that we
have not yet understood (a similar situation is known to arise in a
related context \cite{Aharony:2003sx}, and a similar surprising agreement
to the simple $b=0$ matrix model was recently observed in
\cite{Dumitru:2003hp}).

\begin{figure}
\epsfig{file=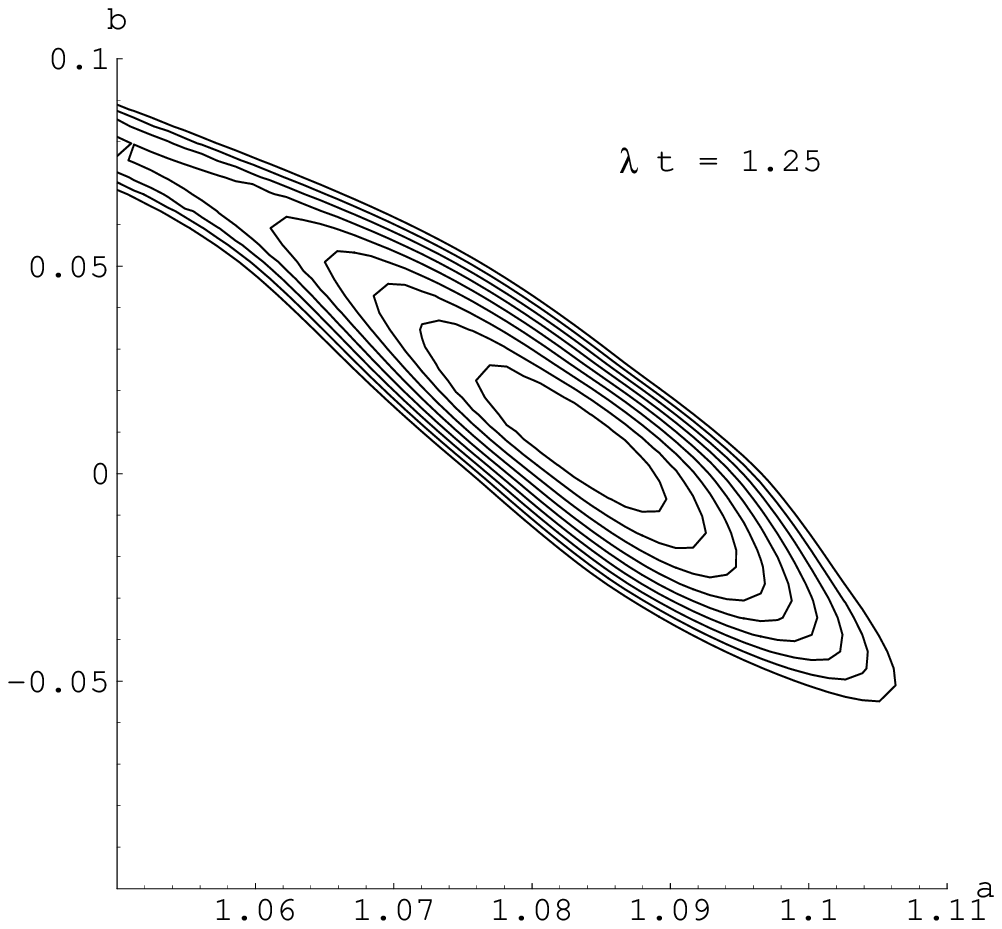,width=3.5in}
\epsfig{file=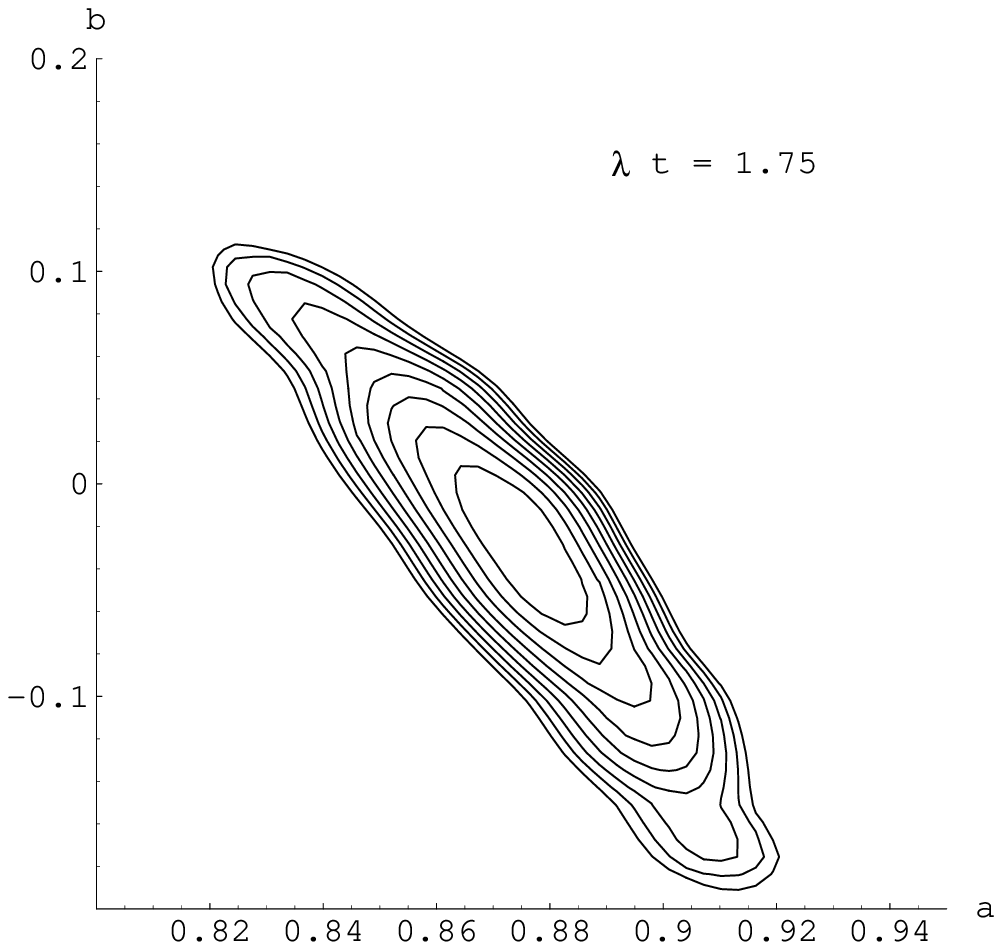,width=3.5in}
\caption{Plots of least squares fit function $L$ in the $(a,b)$ plane,
for values of $\lambda' t$ just below and just above the transition
point. For each plot the contour interval is chosen to equal the
minimum value of $L$, and hence the contours give an indication of
decline of goodness of fit away from the minimum. For both plots we
see the most likely value of $b$ is numerically small.
\label{fig:likely}
}
\end{figure}

We now attempt to quantify the smallness of $b$. Using the lattice
Monte-Carlo data for various $N = 6, 12, 20$, we compare $u_i$ from
the matrix model with the lattice calculation. Note that even for the
best values we do not
expect a perfect fit as our Landau-Ginzburg model can be further
refined by adding in higher order terms in $|\mathrm{Tr}(U)|$, as well
as terms depending on traces of higher powers of $U$. 
We use a least squares fit to test the lattice data $u_i$ and
the Landau-Ginzburg matrix model predictions
$u^{\mathrm{MM}(a,b)}_i$ for particular $(a,b)$,
defining the goodness of fit function $L$ to be minimized as,
\begin{equation}
L(a,b) = \sum_{i=1,2} \sum_{N=6,12,20} \left(
u^{\mathrm{MM}(a,b)}_i(N) - u_i(N) \right)^2.
\end{equation}
In figure \ref{fig:likely} we plot contours of $L$ in the $a,b$ plane,
calculated for two values of $\lambda' t$ on either side of the
transition point. In each case contour intervals are chosen to equal
the minimum value of $L$, and thus indicate goodness of fit. We see
that both plots give a most likely $b$ close to zero.

It is noteworthy that, for any nonzero $b$, the Landau-Ginzburg model
\eqref{eq:lg} possesses exactly 3 saddle points \cite{Aharony:2003sx}
that match perfectly
with the 3 gravitational phases -- the uniform string, the non-uniform
string and the black hole -- that we have described in Section
\ref{sec:dual}.  When $b<0$, the non-uniform string is never
thermodynamically favoured, and figure \ref{fig:phase} qualitatively
captures the thermodynamics of our system. On the other hand, 
when $b>0$, the non-uniform
string phase is the thermodynamically favoured solution throughout its
existence; in fact in such a case it is the (thermodynamic) end point
of the Gregory-Laflamme transition. This is similar to the expected
behaviour of the Gregory-Laflamme transition when $d\geq 14$ 
\cite{Sorkin:2004qq}.

%==============================================================================
%
\section{Summary}
\label{sec:summary}
%
%==============================================================================

Holography predicts that $1+1$-dimensional maximally supersymmetric 
Yang-Mills theory on a circle of circumference $L$
undergoes a first order phase transition at
strong 't Hooft coupling at a critical temperature equal to or larger
than $T_{GL}=
3(2\pi a(0))^2/4\pi^{1/2}L\sqrt{\lambda}$, where $a(0)\approx
0.37$.  The order parameter for this phase transition is the eigenvalue
distribution (uniform at high temperature, clumped at low temperature)
of the Wilson loop along the spatial circle.  
For the dual supergravity theory, the relevant
dynamics is the Gregory-Laflamme instability.

At weak coupling, we have found good evidence for a sharp phase
transition between identical phases at a temperature $T_c \simeq
1.4/\lambda L^3$. We conjecture that a similar transition occurs at
all values of the 't Hooft coupling.  This picture suggests that the
dynamics of compactified horizons probed here, essentially governed by
the Gregory-Laflamme instability, is stable to $\alpha'$ corrections.
Note that a weak coupling behaviour $t_c \sim 1 / \lambda'$ is
replaced at strong coupling by $t_c \sim 1 / \sqrt{\lambda'}$; similar
changes in the dependence on the coupling constant have been observed
in many other cases of theories with holographic duals, for example in
the quark-anti-quark potential in the ${\cal N}=4$ SYM theory
\cite{{Maldacena:1998im},{Rey:1998ik}}.

We discovered that at weak coupling
the behaviour of the theory near the transition is
very well approximated by a simple Landau-Ginzburg model \eqref{eq:lg}.
It is intriguing that \eqref{eq:lg} is also the Landau-Ginzburg model
that governs the dynamics of deconfinement transitions in weakly
coupled Yang-Mills theories \cite{Aharony:2003sx}.  In fact, viewed
from the point of view of the $0+1$ dimensional Euclidean bosonic gauge theory
we analyzed above, the transition we have numerically studied above 
is precisely the deconfinement transition.
Thus, the Gregory-Laflamme and deconfinement phase
transitions appear to lie in the same ``universality class''.

Our Landau-Ginzburg model has precisely 3 distinct phases -- the
uniform, non-uniform and localized phases -- and these agree with
conjectured and known results in the gravity. Thus, it would be
tempting to extrapolate to strong coupling and conclude that these 3
solution branches (including also their reidentifications with larger periods,
for instance a single 
black hole solution suitably scaled and reidentified to give
a solution for $n$ black holes with the same compactification radius)
exhaust the static gravity solutions, and that there are no more
solution branches to be found.  To investigate this from the gravity
side appears to be a very difficult problem.

In order to gain more insight into the SYM phase diagram, it is
obviously important to determine the order of the transition at weak
coupling and to understand why it lies on, or so near to, the cusp 
between first and second order.

While we have focussed on $1+1$ dimensional Yang-Mills in this letter, 
the analysis of
section 2 may trivially be extended to relate the behaviour of
D0-branes on $T^p$ to the thermodynamics of maximally supersymmetric
$p+1$ dimensional Yang Mills on a $p$-torus, at least for $p\leq 3$.
It would be interesting to verify that these gauge theories also
undergo Gregory-Laflamme like phase transitions\footnote{See
\cite{Kiskis:2003rd,Narayanan:2003fc} for related work in pure
Yang-Mills theory.}.

In this note we have studied the theory at finite temperature, and we
have not attempted to address issues concerning the
dynamical end-point of the Gregory-Laflamme type transition that we
have studied \cite{Choptuik:2003qd}. Horowitz and Maeda
\cite{Horowitz_Maeda1,Horowitz_Maeda2} have argued that, within
general relativity, a black string like phase is unable to decay to a
black hole phase in finite time.  In this paper, we have identified
the uniform string with a uniform eigenvalue distribution of a Wilson
line operator; the black hole is a clumped eigenvalue distribution of
the same operator. A cursory analysis reveals no barrier for a
dynamical transition between these two phases. It would certainly be
interesting to understand this better.

%==============================================================================
%
\section*{Acknowledgements}
%
%==============================================================================

We would like to thank R. Emparan, R. Gopakumar, D. Gross,
A. Hashimoto, D. Jafferis, B. Kol, H. Kudoh, I. Mitra, R. Myers,
A. Neitzke, M. Rangamani, H. Reall, B. Sathiapalan, S. Shenker, A. Strominger,
C. Vafa, and X. Yin for useful comments and conversations. We would
especially like to thank J. Maldacena for a discussion that initiated
this project, and K. Papadodimas and M. Van Raamsdonk for numerous
invaluable discussions, as well as for an enjoyable collaboration on a
companion paper \cite{torus}. OA would like to thank Harvard
University for its hospitality during the work on this project. SM
would like to thank the Tata Institute for Fundamental Research and
the organizers of the Indian String Meeting in Kanpur for hospitality
while this work was in progress. TW would like to thank the Perimeter
Institute for hospitality during work on this project.  The work of OA
was supported in part by the Israel-U.S. Binational Science
Foundation, by the ISF Centers of Excellence program, by the European
network HPRN-CT-2000-00122, and by Minerva. OA is the incumbent of the
Joseph and Celia Reskin career development chair. The work of JM was
supported in part by an NSF Graduate Research Fellowship.  The work of
SM was supported in part by DOE grant DE-FG03-91ER40654, by the NSF
career grant PHY-0239626, and by a Sloan fellowship. The work of TW
was supported by the David and Lucille Packard Foundation, Grant
Number 2000-13869A.

%==============================================================================
%
\section*{Appendix A}
\label{sec:appendix}
%
%==============================================================================

In this appendix we demonstrate how to map results about general 10-d static
pure gravity solutions into results about SYM energy and
entropy, by adding charge to the solutions, taking the decoupling limit 
and translating to SYM variables, as discussed in section
\ref{sec:dual}. Once the relevant solutions are computed this shows explicitly
how to map them into predictions for phases of the SYM theory.
Our discussion is phrased in arbitrary dimension, 
even though the correspondence to SYM is only
valid in 10-d. This allows us to examine also the 6-d gravity example
where detailed numerical solutions have been computed, and to make
analogies with our 10-d case of interest.

Let us take a branch of static uncharged 
$d$-dimensional solutions which asymptote
to $R^{d-2,1}\times S^1$. The
metrics can be written in the general form
\begin{equation}
d\tilde{s}^2_{(d)} = - A^2 d\tau^2 + V^2 \left( dr^2 + r^2
d\Omega^2_{d-3} \right) + B^2 dy^2,
\label{eq:staticmetric}
\end{equation}
where $y$ is the compact circle coordinate ($y\equiv y+L$), 
and $A, V, B$ are
functions of $r, y$ which go to unity at large $r$. We are interested
in black solutions for which $A$ vanishes at the horizon. We follow
the 10-d prescription of section \ref{sec:dual} for generating charged
solutions from \eqref{eq:staticmetric} : we add an extra
periodic coordinate $x$, boost in this direction with parameter
$\beta$ and then dimensionally reduce on $x$. This yields a new
Einstein frame metric,
\begin{equation}
ds^2_{(d)} = \phi^{+\frac{2}{d-2}} \left[
  -\frac{A^2}{\left(\cosh^2\beta - A^2 \sinh^2\beta \right)} d\tau^2 +
  V^2 \left( dr^2 + r^2 d\Omega^2_{d-3} \right) + B^2 dy^2\right],
\label{eq:chargedmetric}
\end{equation}
where $\phi$ (the size of the $x$-direction) becomes a non-trivial
dilaton in the supergravity,
\begin{equation}
\phi^2 = \cosh^2\beta - A^2 \sinh^2\beta,
\end{equation}
and now the solution is also charged under a 1-form gauge potential
coming from the $g_{\tau x}$ component of the metric (analogous to the
RR 1-form potential in the 10-d case of section \ref{sec:dual})\footnote{See
also \cite{Bostock:2004mg} for an analogous transform.}.
It is easy to see that the surface gravity at the horizon, and hence the
temperature, is simply redshifted by a factor of $\cosh\beta$. In
addition, since $A$ vanishes at the horizon, the entropy is simply
scaled by this same factor. 

Asymptotically, we can always expand the
metric as
\begin{eqnarray}
A & = & 1 - a(\xi) \left( \frac{L}{r} \right)^{d-4} + \ldots \cr
B & = & 1 + b(\xi) \left( \frac{L}{r} \right)^{d-4} + \ldots
\end{eqnarray}
with $a, b$ essentially giving the two asymptotic charges of the
gravity solution. The dimensionless variable $\xi$ parameterizes a
branch of solutions of the form \eqref{eq:staticmetric} at fixed $L$.
Using $a$ and $b$ we can calculate the ADM mass of the solutions 
\eqref{eq:chargedmetric},
\begin{equation}
M = \frac{L^{d-3}}{16 \pi G} \Omega_{d-3} \left( 2 (d-3) a - 2 b + 2 (d-4) a \sinh^2 \beta \right),
\end{equation}
and their charge
\begin{equation}
Q = \frac{L^{d-3}}{16 \pi G} \Omega_{d-3} \left( (d-4) a \sinh 2 \beta \right).
\end{equation}
The entropy and temperature of \eqref{eq:chargedmetric} are given by
\begin{equation}
S = \frac{L^{d-2}}{4 G} \cosh\beta \Omega_{d-3} s(\xi), \;\;\;
T = \frac{1}{L \cosh\beta} t(\xi),
\end{equation}
for some functions $s, t$ which again depend only on $\xi$ (and not on $\beta$).

Note that for fixed asymptotic circle length $L$, the new solutions
obey the first law of black hole thermodynamics,
\begin{equation}
dM(\xi,\beta) = T(\xi,\beta) dS(\xi,\beta) + \mu(\xi,\beta) dQ(\xi,\beta)
\label{eq:firstlaw}
\end{equation}
with $\mu = \tanh\beta$, provided that the uncharged solution branch
\eqref{eq:staticmetric} obeys
\begin{equation}
dM_0(\xi) = T_0(\xi) dS_0(\xi),
\end{equation}
where $M_0, T_0, S_0$ are the thermodynamic quantities for the uncharged
solution (with $\beta = 0$). Note that in order to show
this, one needs to use the relation
\begin{equation}
\frac{L^{d-3}}{16 \pi G} \Omega_{d-3} \left( 2 (d-4) a \right) = T_0 S_0,
\label{eq:cunningformula}
\end{equation}
which can be derived from \cite{Harmark:2003dg,Kol:2003if}.

Next, we wish to take the near-extremal limit by taking $\beta$ to infinity.
We define the energy above extremality ${\cal E} = M - Q$, and the
`reduced entropy' ${\cal S} = S/\sqrt{Q}$. These quantities remain
finite in the infinite $\beta$ limit, 
even though the quantities $M$, $Q$, and $S$ diverge (for fixed $L, G$).
We may write them as
\begin{equation}
{\cal E}(\beta \rightarrow \infty) = p(\xi) \frac{L^{d-3}}{G}, \;\;\;
{\cal S}(\beta \rightarrow \infty) = q(\xi) \sqrt{\frac{L^{d-1}}{G}}, 
\end{equation}
where the infinite $\beta$ 
limit defines the quantities $p(\xi), q(\xi)$, giving
\begin{equation}
p(\xi) = \frac{\Omega_{d-3}}{16 \pi} \left( (d-2) a(\xi) - 2 b(\xi) 
\right),\;\;\;
q(\xi) = \sqrt{ \pi \Omega_{d-3} } \frac{s(\xi)}{\sqrt{ 2 (d-4) a(\xi)}}.
\end{equation}

As in M(atrix) theory \cite{Banks:1997vh}, in the 10-d case 
the infinite boost limit is equivalent to a decoupled SYM theory, 
and in fact the infinite boost limit is
closely related to the near-horizon decoupling limit discussed in section
\ref{sec:dual} (as explained most clearly in
\cite{Polchinski:1999br}).  After translating the quantities above to
the SYM theory one finds that the SYM (dimensionless) energy $\epsilon$ and
entropy $\sigma$ are simply functionally related to $p$ and $q$ by
\begin{equation}
\epsilon \propto \frac{N^2}{\lambda'^2} p(\xi), \;\;\; 
\sigma \propto \frac{N^2}{\lambda'^{3/2}} q(\xi).
\label{eq:translation}
\end{equation}
Since the factors of $N$ and $\lambda'$ are the same for all solutions
of the form \eqref{eq:staticmetric}, this shows that in order to
analyze which phase is preferred in the SYM theory, we do not have to
discuss the full decoupling limit of the near extremal solutions, but
instead we can just consider the quantities ${\cal E, S}$. Their
construction from $M, S$ and $Q$ gives the functions $p(\xi), q(\xi)$
in the simpler infinite $\beta$ limit, and this is related by
\eqref{eq:translation} to the behaviour of the SYM energy and entropy
(as functions of $\xi$) in the actual decoupling limit.

As an example of the formalism of this section, 
the $d$-dimensional uncharged uniform black string metric is
\begin{equation}
ds^2= -(1-\left({r_0 \over r}\right)^{d-4} ) d\tau^2 + {dr^2\over {1
    -\left({r_0 \over r}\right)^{d-4}}} + r^2 d\Omega_{d-3}^2 + dy^2.
\label{eq:metstring}
\end{equation}
Putting this in the form of \eqref{eq:staticmetric} and applying the
transformation to the near extremal charged string gives
\begin{equation}
p(\xi) = \frac{\Omega_{d-3}}{16 \pi} \frac{d-2}{2} \xi^{d-4}, \qquad 
q(\xi) = \sqrt{ \frac{\pi \Omega_{d-3}}{d-4} \xi^{d-2} },
\label{eq:solpq}
\end{equation}
where for this branch we have defined $\xi = r_0/L$. For the
localized solution with $r_0 \ll L$ we may simply approximate the
unboosted starting metric by the $d$-dimensional Schwarzschild
solution, and in this limit we find that $p$ and $q$ are given by similar
expressions to \eqref{eq:solpq} but with $d$ replaced by $d+1$.

\begin{figure}
\epsfig{file=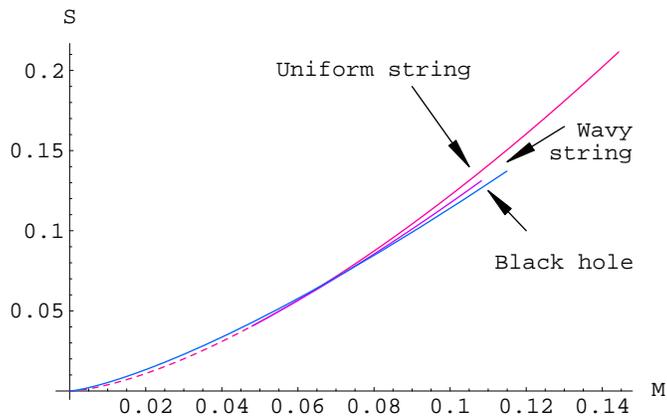,width=3.5in}
\caption{
\label{fig:unboosted6d}
Plot of entropy as a function of mass for the static uncharged black hole,
non-uniform and uniform strings in $d=6$.  The results are taken from
\cite{Kudoh:2003ki}, where the black hole branch is only partially
constructed, and over this range behaves very similarly to a 6-d
Schwarzschild solution. Recent calculations indicate that the black
hole branch turns around and connects to the non-uniform string branch
\cite{Toby_Kudoh}. }
\end{figure}

In 6-d numerical solutions for the uncharged non-uniform string and
black hole branch on a compact circle have been found. In figure
\ref{fig:unboosted6d} we plot the entropy as a function of mass for
these solutions, with the $y$ circle having unit length. We may then
ask whether this diagram remains qualitatively similar when charge is
added as discussed above, and the near-extremal limit is taken. Of
course, since these solutions are not in 10-d we cannot use these
transformed near extremal solutions to directly make statements about
strongly coupled YM theory.

\begin{figure}
\epsfig{file=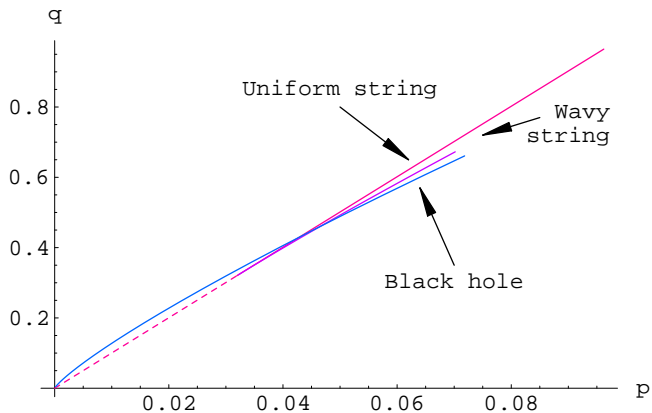,width=3.5in}
\caption{
\label{fig:boosted6d}
Plot of $p, q$ for the near-extremal solutions obtained by transforming the 
6-d uncharged static solutions.
In $d=10$ this would give the relevant backgrounds dual
to SYM. Here in $d=6$ there is no dual, but we think of the functions
$p,q$ as analogous to energy and entropy, as in the $d=10$ case. Note
then that the entropy-mass ordering of the solutions remains the same
as for the uncharged solutions of figure \ref{fig:unboosted6d}, 
even though a priori this is not
guaranteed.  }
\end{figure}

In figure \ref{fig:boosted6d} we plot $q$ as a function of $p$ for the same
solutions (in the 10-d case this would give $\sigma$ and $\epsilon$ in the
SYM).  We see that the ordering of the solutions appears to be
preserved when taking this near extremal limit transform of the
original uncharged solutions. We expect a similar result to hold also
for 10-d, where the solutions can be related to YM phases.
In the following appendix we show that in 10-d near the point where
the non-uniform solutions emerge from the uniform branch, the same
qualitative behaviour is indeed found in the near-extremal charged
limit as for these 6-d solutions.

%==============================================================================
%
\section*{Appendix B}
\label{sec:appendix2}
%
%==============================================================================

The behaviour of static uncharged non-uniform strings near the
critical uniform string solution can be computed using Gubser's method
\cite{Gubser}. Results in 10-d were recently obtained by Sorkin
\cite{Sorkin:2004qq}, who showed that for $d \leq 13$ the non-uniform
strings have lower entropy than the uniform ones of the same mass.
We now repeat these calculations, and then use the results from the
previous appendix in order to obtain the behaviour of the 10-d near
extremal non-uniform charged solutions relevant for this paper.

Firstly, repeating Sorkin's analysis and using his notation for
brevity, perturbing to third order about a unit horizon radius ($r_0=1$)
uncharged uniform string we find
\begin{eqnarray}
k_0  & = & \;\;\, 2.30, \qquad k_1 = 0.\cr
dT/T & = &      - 0.85 \lambda^2 + O(\lambda^4), \qquad A_{inf} = -4.19, \cr
dS/S & = & \;\;\, 9.28 \lambda^2 + O(\lambda^4), \qquad B_{inf} =  1.66,
\end{eqnarray}
where $dT/T, dS/S$ give the fractional difference of $T, S$ for a
non-uniform string with non-uniformness $\lambda$ (which we can take
to define $\xi$ for this branch) compared to the critical uniform
string $\lambda = 0$. In Sorkin's notation $A_{inf}, B_{inf}$ give the
asymptotics and can be easily related to our $a, b$. As stated above,
for these values we have used the gauge condition $k_1 = 0$. This
yields
\begin{equation}
\frac{S_{\mathrm{non-uniform}}}{S_{\mathrm{uniform}}} = 1 - 2.2 \lambda^4 + O(\lambda^6),
\label{eq:statvaccase}
\end{equation}
where both solutions have the same mass, and hence at least near the
critical uniform solution, as in 6-d, the uncharged uniform strings
have higher entropy than the non-uniform strings.

Next, we wish to understand this behaviour for the charged limit
outlined in the previous appendix. Again we consider the gravitational
quantities ${\cal E}, {\cal S}$ which, as in the previous appendix,
are simply proportional to the YM $\epsilon, \sigma$ in the infinite
$\beta$ limit,
\begin{equation}
{\cal E} = M - Q, \qquad
{\cal S} = \frac{1}{\sqrt{Q}} S, \qquad
{\cal T} = \sqrt{Q} T,
\end{equation}
and we also define ${\cal T}$, the `reduced' temperature, which again
is finite in the infinite $\beta$ limit, where it is proportional to
the YM temperature.

From now on all quantities are given for $\beta \rightarrow \infty$.
We may then use our previous appendix results to find
\begin{eqnarray}
d {\cal E} / {\cal E} & = & 7.6 \lambda^2  + O(\lambda^4), \cr
d {\cal S} / {\cal S} & = & 5.1 \lambda^2  + O(\lambda^4), \cr
d {\cal T} / {\cal T} & = & 3.3 \lambda^2  + O(\lambda^4),
\end{eqnarray}
where again these denote the fractional difference from the critical
uniform string. For infinite $\beta$, the first law
\eqref{eq:firstlaw} implies that
\begin{equation}
d {\cal E} = {\cal T} d {\cal S}
\end{equation}
(using equation \eqref{eq:cunningformula}). Since the uniform and
non-uniform branches merge at the marginal solution, and hence have
the same ${\cal T}$ there, the ratio ${\cal
S}_{\mathrm{non-uniform}}/{\cal S}_{\mathrm{uniform}}$, for the same
energy ${\cal E}$, must also go as $\lambda^4$, with the first law
implying that the potential leading $\lambda^2$ term
vanishes. However, as Gubser showed, fortunately we may use the first
law to compute this difference without resorting to higher order
perturbation theory. In fact, we find
\begin{equation}
\frac{{\cal S}_{\mathrm{non-uniform}}}{{\cal S}_{\mathrm{uniform}}} =
1 - 2.0 \lambda^4 + O(\lambda^6),
\end{equation}
where the solutions have the same `reduced' energy ${\cal E}$.

Thus, finally, we confirm that as for the static uncharged case 
\eqref{eq:statvaccase}, also in
the $\beta \rightarrow \infty$ limit the 10-d non-uniform solutions
have lower entropy ${\cal S}$ for the same energy ${\cal E}$. Since (by
design) in this limit these quantities are proportional to the dual YM
entropy and energy, this predicts that the same is true in the YM
theory at strong coupling.

As we are interested in YM thermodynamics in which we fix the temperature
rather than the energy, we may simply extend this calculation to
compute
\begin{equation}
\frac{{\cal F}_{\mathrm{non-uniform}}}{{\cal F}_{\mathrm{uniform}}} =
1 - 8.0 \lambda^4 + O(\lambda^6),
\end{equation}
where ${\cal F} = {\cal E} - {\cal T} {\cal S}$ is the `free energy'
of the solution (which is again proportional to the YM free energy), and
the two solutions which are compared 
have the same ${\cal T}$. This shows that the
non-uniform phase has a higher free energy than the uniform one
(since ${\cal F}_{\mathrm{uniform}}$ is negative) at fixed temperature
near the marginal point, and is thus thermodynamically disfavoured. We
use this information to aid in our construction of a likely phase
diagram for the YM theory in figure \ref{fig:phase}, and to argue that
the strongly coupled YM theory goes through a first order phase transition.

%==============================================================================
%
\section*{Appendix C}
\label{sec:appendixc}
%
%==============================================================================

The system we have investigated in this paper -- $1+1$ dimensional
maximally supersymmetric Yang-Mills theory on a circle -- has been
previously studied in different contexts. In this appendix we describe
how our results fit in with those of previous investigations.

\subsection{Reduction to the symmetric product CFT at very strong coupling}

In the limit of very strong coupling, $\lambda' \gg N^2$, the SYM
theory \eqref{eq:ymaction} flows, at low enough energies, to a $1+1$
dimensional superconformal field theory which is a sigma model on the
target space $(R^8)^N/S_N$ where $S_N$ is the permutation group on $N$
objects
\cite{Harvey:1995tg,Bershadsky:1995vm,Motl:1997th,Dijkgraaf:1997vv,Itzhaki:1998dd}
\footnote{The condition $\lambda' \gg N^2$ is required to flow
to the orbifold CFT at energies or temperatures of order $1/L$. Outside
the 't Hooft limit one is sometimes interested in energies or temperatures
scaling as $1/NL$, and for these to be well-described by the orbifold CFT
it is enough to require $\lambda' \gg 1$.}.
The fermions in \eqref{eq:ymaction} are periodic around the spatial
circle; as a consequence, the CFT is in the Ramond sector, and hence
is in the `long string phase' \cite{Dijkgraaf:2000fq} at any nonzero
temperature.  This implies that, in the parent gauge theory, the
spatial holonomy is the shift matrix, whose eigenvalues are uniformly
distributed over the circle. This is consistent with our results; as
we have described (see \eqref{eq:tcrit}), at strong coupling in the
't Hooft limit \eqref{eq:ymaction} is in
the uniform string phase for $t \gg {1 \over \sqrt{\lambda'}}$,
i.e. at every temperature in the limit $\lambda' \to \infty$
considered here.

If we instead study \eqref{eq:ymaction} with fermions antiperiodic
around the spatial circle, then it would flow, at very strong
coupling, to the symmetric product CFT in the Neveu-Schwarz
sector. This CFT, in the large $N$ limit, undergoes a phase transition
between a low temperature short string phase and a high temperature
long string phase at exactly $t=1$
\cite{Aharony:1999ti,Dijkgraaf:2000fq,David:2002wn}. In the short
string phase, the spatial holonomy of the parent gauge theory is
peaked around the identity matrix so this corresponds to our black
hole phase.  We thus conclude that \eqref{eq:ymaction} with
antiperiodic fermions and at very strong coupling undergoes a first
order black hole-black string transition at $t=1$. This is consistent
with results to appear \cite{torus}.

\subsection{Relation to Matrix String Theory}

Another interesting limit of the SYM theory \eqref{eq:ymaction} is the
limit of large $N$ with constant $g_{YM}$, which appears in Matrix string
theory \cite{{Banks:1997vh},{Motl:1997th},{Dijkgraaf:1997vv}}. 
When we lift the background \eqref{eq:sugram}
to M theory as described in section \ref{sec:uniform}, it is interpreted
as the near-horizon limit of an M theory configuration (with the eleven
dimensional Planck scale $M_{11} \propto (L / g_{YM}^2)^{1/3} / \alpha'$)
where we have $N$ units of
momentum on the circle of the eleventh dimension, smeared along a
transverse circle (say, the ninth dimension). If we now reduce this
background back to type IIA string theory along the ninth dimension rather
than along the eleventh dimension, we find a type IIA configuration with
a constant dilaton (with the string coupling proportional to
$1 / g_{YM} L$, and the string tension proportional to 
$1 / (g_{YM} \alpha')^2$), and still involving $N$ units of momentum around a
compact circle, but now with no additional circles. As described, for
instance, in \cite{Polchinski:1999br}, the near-horizon limit of this
configuration can be identified with the DLCQ limit (or the infinite
momentum frame) which is relevant to Matrix string theory.

Note that, as in the previous subsection, in the matrix string limit we
have $\lambda' \to \infty$, and then (for any finite temperature in the
large $N$ limit) \eqref{eq:ymaction} is always in the uniform
string phase and undergoes no transitions\footnote{For a discussion of
matrix string thermodynamics, see for example
\cite{Sathiapalan:1998gh}.}.

%==============================================================================
%
%\newpage 
\bibliography{letter}
%
%==============================================================================

%==============================================================================
%
\end{document}